# Characterization of a Superconducting Microstrip Single-Photon Detector Shunted with an External Resistor

**Yu-Ze Wang[1,2,3], Wei-Jun Zhang[1,2,3*], Guang-Zhao Xu[1,2,3], Jia-Min Xiong[1,2,3], Dong-Hui Fan[1,2,3], Zhi-Gang Chen[1,2,3], Xing-Yu Zhang[1,2], Zhen Wang[1,2,3], and Li-Xing You[1,2,3]**

[1] State Key Lab of Functional Materials for Informatics, Shanghai Institute of Microsystem and Information Technology (SIMIT), Chinese Academy of Sciences (CAS), 865 Changning Rd., Shanghai, 200050, China.
[2] CAS Center for Excellence in Superconducting Electronics (CENSE), 865 Changning Rd., Shanghai, 200050, China.
[3] Center of Materials Science and Optoelectronics Engineering, University of Chinese Academy of Sciences, Beijing 100049, China.

[*]E-mail: zhangweijun@mail.sim.ac.cn

## Abstract

A superconducting microstrip single-photon detector (SMSPD) generally requires a shunt resistor to avoid latching, caused by its high current-carrying capacity and low kinetic inductance. Here, the effect of the shunt resistor on the behaviors of microbridge SMSPDs was investigated. We analyzed the change in equivalent switching current at different shunt resistances in two ways and determined the operating current range using intrinsic dark count rate (iDCR) curves. We observed that the reduction in shunt resistance can increase the operating current range, which helps to improve the internal detection efficiency (IDE) and reduce the iDCR. However, the reduction in the shunt resistance can reduce the pulse amplitude and increase the pulse decay time, which can degrade the timing jitter and count rate performance of the SMSPD. The trends of the experimental results can be qualitatively reproduced using a circuit model for an SMSPD with a shunt resistor, which provides useful information for the selection of shunt resistors. Furthermore, we report the improved detection performance of a helium-ion-irradiated SMSPD shunted with a small resistance of 5.2 Ω. We observed a weak IDE saturation with a bias current at a wavelength up to 2000 nm and a nonlinear relation between detection current and photon energy.

## 1. Introduction

Recently, cutting-edge scientific fields, such as deep-space laser communication [1], satellite-to-earth optical quantum key distribution [2], confocal imaging [3], and dark-matter detection [4], have achieved rapid development. Urgent application requirements have been proposed for single-photon detectors with a large active area, high speed, and low timing jitter. Traditional semiconductor single-photon detectors, such as avalanche photodiodes (APDs), have an active area of tens to hundreds of microns in diameter (e.g., ~ 100 μm) but exhibit low system detection efficiency (SDE, ~35% at 1310 nm) and high dark count rate (DCR, ~4.5 × $10^5$ cps) in the near-infrared region [5], which limit their application range. Superconducting nanowire single-photon detectors (SNSPDs) exhibit excellent performance, such as a high SDE > 90%, low DCR < 0.1 cps, low time jitter < 3 ps, high maximum count rate exceeding 1.5 Gcps, and broadband sensitivity from 300 nm to 10 μm [6-11]. However, the performance of SNSPDs with large active areas (diameter > 50 μm) still requires improvement in terms of speed and timing jitter. Compared with SNSPDs, recently developed superconducting microstrip single-photon detectors (SMSPDs) [12-14] exhibit micron-scale strip width, lower kinetic inductance ($L_K$), and higher switching current ($I_{sw}$), and are easier for fabricating large-active-area detectors. SMSPDs have received extensive attention in the application field of large-active-area detectors. The first prerequisite for realizing high-efficiency SMSPDs is a saturated internal detection efficiency (IDE), and a saturation detection plateau implies an IDE of ~ 100%. In 2020, SMSPDs made of amorphous materials, such as ultrathin WSi and MoSi, were reported to achieve a saturated detection of 1550 nm photons [15, 16]. In 2021, Xu *et al.* reported that an NbN SMSPD irradiated using helium ions achieved near-saturation detection at a wavelength of 1550 nm, with an SDE exceeding 90% [17]. Furthermore, numerous novel findings were raised in the field of SMSPDs, such as those related to materials [18], geometric layout design [19, 20], and optical lithography fabrication [18, 21, 22].

The IDE of SNSPDs and SMSPDs essentially depends on the transition probability of the strip cross-section from the superconducting state to the resistive state. Conventional SNSPDs harness the hotspot effect through the fabrication of narrow strips (typically 30–200 nm) [23] because the size of the photon-induced hotspot that initiates the transition is typically in the order of tens of nanometers. In contrast, theoretically, if the operating current in the strip can be biased sufficiently close to the depairing current ($I_{dep}$) of the superconducting Cooper pair, for example, > $0.7I_{dep}$, the heat generated by the vortex or antivortex motion driven by the current can be sufficient for creating a resistive state cross-section [12, 13]. Under this condition, the detection of photons can be independent of the strip width, as long as the width is less than the Pearl length. Therefore, making the operating current as close as possible to the $I_{dep}$ is a key factor for realizing high IDE in SMSPDs. However, owing to the high current-carrying capacity and low $L_K$ characteristics of the typical SMSPDs, a shunt resistor is usually required to avoid latching [14, 17, 24, 25]. With the shunt resistor, the SMSPD can be subjected to a bias current closer to its critical current since the heat generation in the strip is reduced, thereby enabling the self-recovery of the strip (returned from the resistive state to the superconducting state).

Shunt resistors have been previously connected in parallel with SNSPDs to prevent latching either at room (~293 K) [26, 27] or cryogenics temperature [28]. In 2012, Brenner *et al.* theoretically and experimentally studied the dynamic mechanism of superconducting ultranarrow nanowires (12–18 nm) in the presence of a shunt resistor [29]. They observed that the value of the shunt resistance strongly affected the statistics of the switching and retrapping currents. They indicated that $I_{sw}$ can be controllably driven very near the $I_{dep}$ through the shunt resistor. For the superconducting microstrip, whose width is considerably larger than the superconducting coherence length, its dissipation is dominated by vortex and antivortex dynamics [24]. The presence of the shunt resistor in SMSPDs reduces the actual current flowing in the strip, thereby slowing down the vortex motion and potentially increasing the switching current of the device. In 2020, Dryazgov *et al.* added a shunt resistor to a phenomenological electrothermal model for the SMSPD and compared the theoretical values of the shunt resistance with the experimental values [30]. In 2021, Korneeva *et al.* studied the influence of shunt resistance on the latching behavior of superconducting strips with different widths, and they observed that the theoretically calculated maximum shunt resistance had an exponential relation with $I_{dep}$ [25]. However, several questions on SMSPDs need to be addressed. First, it is still unclear if there is an "ideal" or "optimal" shunt resistance for SMSPDs, such that the current can be biased as high as possible close to the $I_{dep}$ of the detector. Second, the effect of shunt resistance on detector behavior, such as the photon response and dark count of SMSPDs, has not been studied. Third, the photon-response behavior of SMSPDs over a broad wavelength range has not been fully understood; this part of the study will reveal the detection mechanism and provide a new method for realizing detectors for applications with a long wavelength beyond 1550 nm [3, 31].

Here, we systematically investigated the effect of shunt resistance on the behavior of SMSPDs. We analyzed the variation in the equivalent switching current at different shunt resistances in two ways and studied the operating current range through intrinsic DCR (iDCR) curves. We qualitatively explained the experimental results through the circuit model of an SMSPD shunt with a resistor and verified the results with the LTspice simulation. To demonstrate the improvement in IDE with a shunt resistor, we characterized the detection performance of a double-spiral SMSPD with a large active area that was irradiated with helium ions and shunted with different resistors. We examined the photon-response behavior of this SMSPD shunted with a 5.2 Ω resistor within a wavelength range of 400–2000 nm. We observed that the SMSPD exhibited weak IDE saturation at a wavelength up to 2000 nm and observed a nonlinear relation between detection current and photon energy.

## 2. Methods and measurements

Here, NbN thin films were deposited by applying direct current (DC) reactive magnetron sputtering to a niobium target in an argon and nitrogen atmosphere. Here, NbN thin films were deposited by direct current reactive magnetron sputtering to a niobium target in mixed argon and nitrogen atmosphere with a background pressure of $\sim 1.3\times10^{-5}$ Pa. The flow rates of Ar and $N_2$ were 30 sccm and 4 sccm, respectively, with a total pressure of 0.27 Pa. The sputtering power is 619 W, with a current of 2.2 A. The film was deposited at ambient substrate temperature. The thickness of the thin films was approximately 7 nm, estimated using the deposition time and rate. After deposition, the NbN-coated wafers were loaded into a 300-mm medium-current ion implanter (Nissin, Exceed 2300RD) and irradiated in different batches with different doses of helium ions with an energy of 20 keV at room temperature [32]. The doses of ion irradiation were $1 \times 10^{16}$ and $5 \times 10^{16}$ ions/cm$^2$. Electron beam lithography and reactive-ion etching (RIE) were employed to fabricate a design pattern of microstrips on the irradiated NbN film. The devices were etched in a $CF_4$ plasma, with a flow rate of 30 sccm, an etching power of 50 W, and an etching time of ~37 s. Finally, coplanar waveguide electrodes were fabricated by ultraviolet (UV) lithography and RIE.

For the experiments, we designed and fabricated two types of devices. The first was a microbridge device with a bridge width of 1 or 2 μm and a length of 10 μm (Figure 1(a)). It was connected to an NbN meander inductor in series, with a wide strip width of 4 μm (Figure 1(b)). The second was a double-spiral strip device (Figure 1(c)) with a strip width of ~ 1 μm, a fill factor of 0.8, and an active area diameter of ~ 50 μm. Microbridge and spiral-strip devices with a bridge/strip width of 1 μm were fabricated using NbN thin films at a high ion irradiation dose ($\sim 5 \times 10^{16}$ ions/cm$^2$), and 2-μm microbridge devices were fabricated using NbN thin films at low ion irradiation doses (~$1 \times 10^{16}$ ions/cm$^2$).

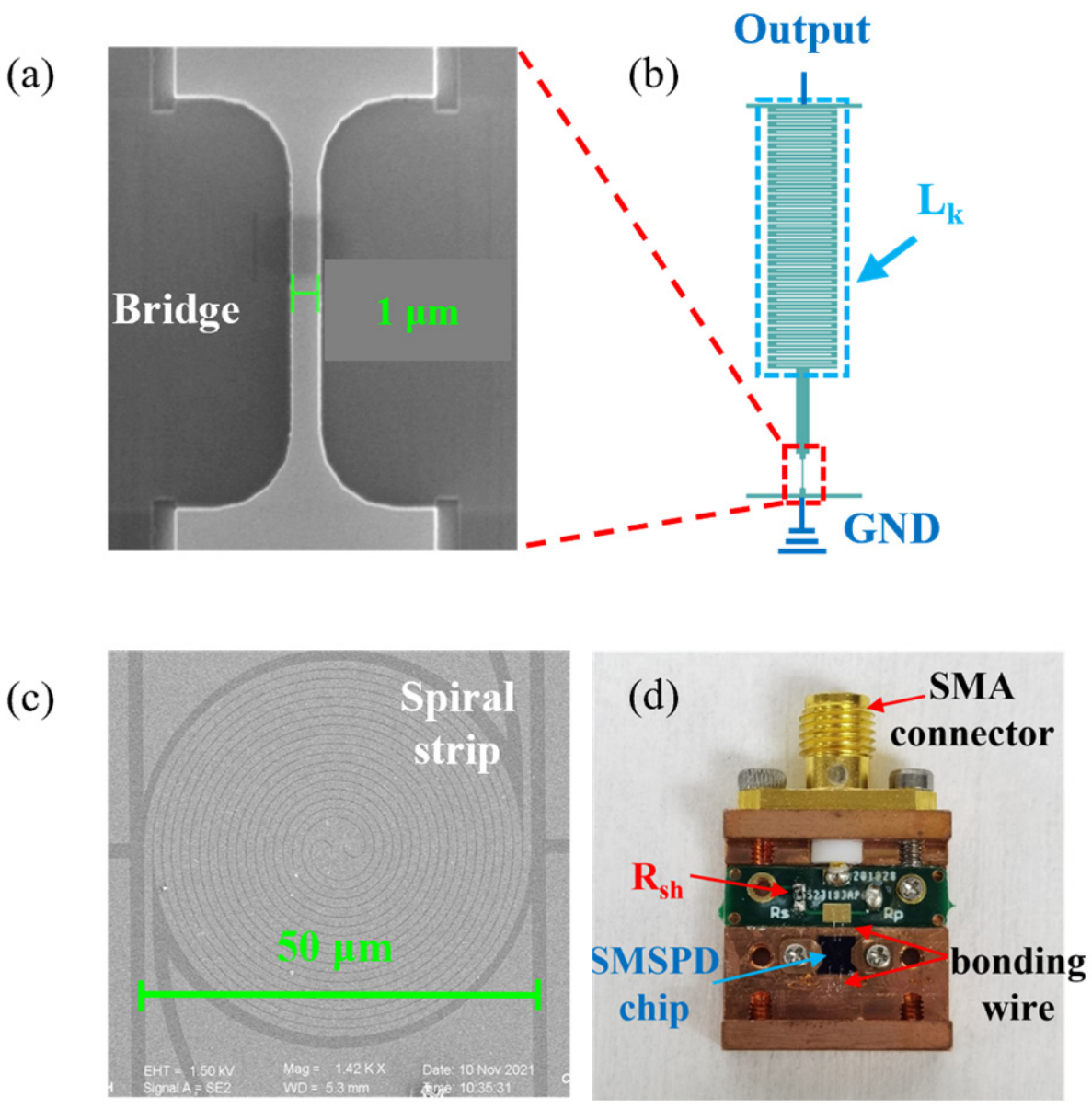

**Figure 1.** (a) SEM image of a 1-μm bridge device. (b) Schematic diagram of the microbridge and a series inductor $L_K$, comprising the NbN thin film. (c) SEM image of a 1-μm spiral strip. (d) Image of an SMSPD package block, where the SMSPD, and a shunt resistor, $R_{sh}$, were placed in the same block.

The fabricated device was glued in the groove of the copper packaging block using varnish (Figure 1(d)). The electrodes of the device were wire-bonded to the gold pad of the printed circuit board (PCB) or the grounded packaging block. A shunt resistor was connected in parallel to the device on the PCB. The $R_p$ pads on the PCB were directly connected with solder at a resistance of approximately 0.1 Ω at room temperature. Here, we characterized the devices at two cryostats: (1) a compact closed-cycle Gifford McMahon refrigeration system with a base temperature of 2.2 K and (2) an adsorption refrigeration system with a base temperature of 0.85 K. The packaged device, coupled with SM28e fiber, was installed on the cold plate of the refrigeration system.

The resistance of the shunt resistor was measured using a multimeter at room temperature (~293 K). The resistance at 2.2 K will slightly increase to approximately 1.2 times that at room temperature. A conventional alternating current (AC)-coupled readout circuit scheme was used. The bias current was provided by a DC voltage source (SIM928, SRS Inc.) in series with a 10 kΩ resistor. This current was fed to the detector through the DC port of a bias tee (ZX85-12G-S+, Mini-Circuits Inc.). The voltage pulse generated by the devices was transmitted to a 50-dB room-temperature low-noise amplifier (LNA-650, RF Bay Inc.) through the AC port of the bias tee and a cryogenic coaxial cable. The amplified pulse signal was read out using a pulse counter (SR400, SRS Inc.) or a high-speed oscilloscope at room temperature.

## 3. Results

### *3.1 Dependence of current–voltage characteristics on shunt resistance*

Figure 2 shows the sweeping current–voltage (I–V) curves of a 2-μm microbridge device with and without a shunt resistor. The measured voltage was extremely low (<5 mV) when the device was connected to a shunt resistor. To distinguish the curves with different shunt resistances, the horizontal axis in Figure 2(a) was plotted on a logarithmic scale. When there was no shunt resistor (i.e., $R_{sh} = \infty$, pink line), the $I_{sw}$ and the retrapping current ($I_r$) of the 2-μm microbridge device were 142.0 and 21.0 μA, respectively. With a decrease in the shunt resistance, the voltage dividing ratio of the device continuously reduced in the entire circuit. As a result, the I–V curve slightly moved to the left (low-voltage region). Figure 2(b) shows an enlarged view of the switching current region, where the horizontal axis is plotted on a linear scale. Taking the I–V curve of $R_s$ = 10 Ω as an example, as the voltage increased, the curve presented three Regions (I, II, III) with different slopes. Region I corresponded to the voltage range (0.4–1.2 mV, not fully shown), where the slope of the I–V curve corresponded to a contact resistance of ~ 5 Ω, and the SMSPD was superconducting. Thus, this part of the resistance was mainly contributed by the resistance of the series bonding wire and the PCB wiring. Region II (1.2–1.5 mV) is a plateau region of the retrapping current with a shunt (denoted by $I_{r\text{-}sh}$), corresponding to the slow growth of the resistive region (or hotspot) in a steady state. Region III (1.5–2 mV) corresponded to the behavior of the device entering the resistive state. At this moment, the device resistance was significantly higher than the shunt resistance; therefore, the circuit impedance was mainly affected by the shunt resistance.

For the cases shunt with a resistor, we defined an equivalent switching current, $I_{sw\text{-}sh}$, at which the current begins to convert from Regions I into II with an increase in the voltage, referred to as the "conventional method" or "determined from the I–V curve." For instance, $I_{sw\text{-}sh}$ ~ 157 μA, at $R_{sh}$ = 10 Ω. As the shunt resistance continuously reduced, the $I_{r\text{-}sh}$ increased, and the plateau of Region II gradually decreased or even disappeared (e.g., $R_{sh}$ = 2 Ω). However, the slope of Region III gradually increased, making the slope switching between Regions I and III difficult to distinguish. Therefore, it is difficult to determine the switching current using the conventional method when the shunt resistance is low.

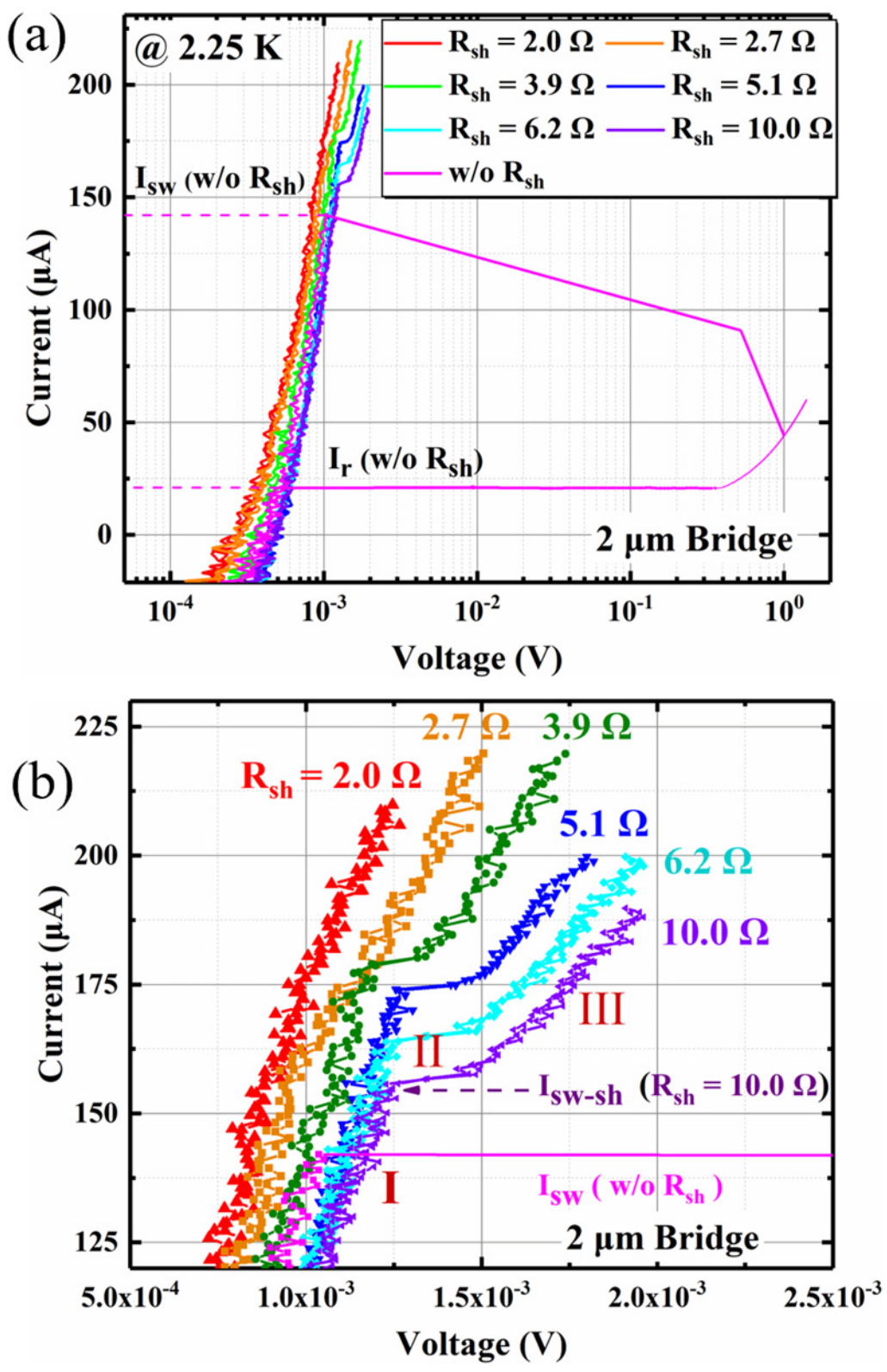

**Figure 2.** (a) I–V characteristics of the 2-μm bridge without (magenta line) and with different $R_{sh}$ values (other colored lines), where the horizontal axis is plotted on a logarithmic scale. (b) Zoomed-in view of the I–V characteristic near the switching-current region but with the horizontal axis on a linear scale. As the voltage increases, the slope of the curve presents three Regions: I, II, and III.

Note that the I-V curve without shunt resistance shows a deviation from those with a shunt. The reason is possibly due to a slight difference in the readout circuits of the cryostats that we used to measure the two sets of data. Specifically, we carried out the I-V curve measurements in two cryostat systems A and B, both of which had a based temperature of ~2.2 K. We installed the chip-mounted block without a shunt into the cryostat system A, whose cryogenic coaxial cables had an average resistance of 4.4 Ω. While the I-V curves with different shunt resistances for the same devices were measured in cryostat system B, which has an average resistance of 3.3 Ω for its cryogenic coaxial cables. Due to the about 1-Ω difference in the readout circuits, the I-V curve without the shunt shifted to the right by $1.5 \times 10^{-4}$ V along the X-axis. This small shift does not affect our conclusions about the shunt resistance dependence of the I-V curves, but it reminds us to pay attention to the small difference in the readout circuits especially when a shunt resistance is included.

### *3.2 Dependence of the intrinsic dark count rate and operating current range on shunt resistance*

Generally, when a shunt resistance is connected, the $I_{sw\text{-}sh}$ extracted from the I–V curve is higher than the $I_{sw}$ measured without a shunt. However, the nominal value of $I_{sw\text{-}sh}$ is not equivalent to an actual increase in the $I_{sw}$ of an SMSPD. The actual increase needs to be determined according to the operating bias current range in photon detection or DCR measurement. During the experiments, we measured the iDCR as a function of bias current ($I_b$) for microbridge devices of different strip widths at different shunt resistances. Here, the iDCR refers to the DCR measured when the device was shielded using aluminum tape without coupling the fiber.

Figure 3(a) shows the iDCR results obtained from the 1-μm bridge device. Dissimilar to the conventional nanowires without a shunt [33], the iDCR behavior of microwires exhibited a nonlinear growth on a semilogarithmic scale with a linear increase in current. Moreover, when the shunt resistance was high (e.g., $R_{sh} \geq 15$ Ω), a "sudden jump" or "inflection point" was observed in the iDCR curve. After the inflection point, the iDCR exhibited a slowly rising plateau. With a decrease in the shunt resistance, the iDCR in the plateau decreased, and the inflection point gradually disappeared, as indicated by the red arrow in Figure 3(a).

It is difficult to discriminate the inflection point in iDCR curves when the device is shunted with a small resistor. Thus, according to the iDCR curve, we used the oscilloscope to monitor the pulse waveforms generated by the device in different operating current regions. This method was referred to as "determined from the iDCR curve." When the device was biased in the normal operating region, the pulses randomly appeared (Figure 3(b-1)). As the bias current increased, the pulses significantly increased and gradually overlapped, after which the device entered the plateau or a relaxation oscillation region (Figure 3(b-4)). The relaxation oscillation of the pulse waveforms was caused by the frequent switching between the superconducting and resistive states, where the electrothermal feedback in the SMSPD was unstable under the interaction between the current-activated hotspot and the kinetic inductance of the SMSPD. Similar behaviors have been observed in SNSPDs[27, 34, 35] . From the different behaviors of the pulse waveform, we identified the current at which the device entered the oscillation region from the normal operating region. We defined this current as the equivalent switching current, $I_{sw-sh}^{DCR}$, determined from the iDCR curve. This method addresses the problem of the determination of the switching current when the switching region is not evident in the I–V or iDCR curve.

Regarding the problem of finding the optimal shunt resistance value, for the device at $R_{sh} \leq 6.8$ Ω, the iDCR (Figure 3(a)) smoothly increased with an increase in the bias current. This phenomenon indicated that the shunt resistor stabilized the vortex dynamics in the strip. Empirically, we believe that a shunt resistance ≤ 6.8 Ω, in this case, was suitable. However, since the shunt resistance would also affect other performances of the device, which are subsequently discussed, the optimal value can be a compromise result. Thus, we choose to use the word "suitable" or "proper." To compare the influence of the shunt resistor on the IDE of the SMSPD, we defined the operating bias current range ($\Delta I_b$) of the SMSPD (Figure 3(a)), i.e., $\Delta I_b = I_{sw-sh}^{DCR} - I_b^*$. Here, $I_b^*$ was the bias current at which the iDCR was equal to 10 cps at a certain $R_{sh}$. Generally, the larger the $\Delta I_b$, the higher the bias current ratio to $I_{dep}$, i.e., a high IDE was expected of the device.

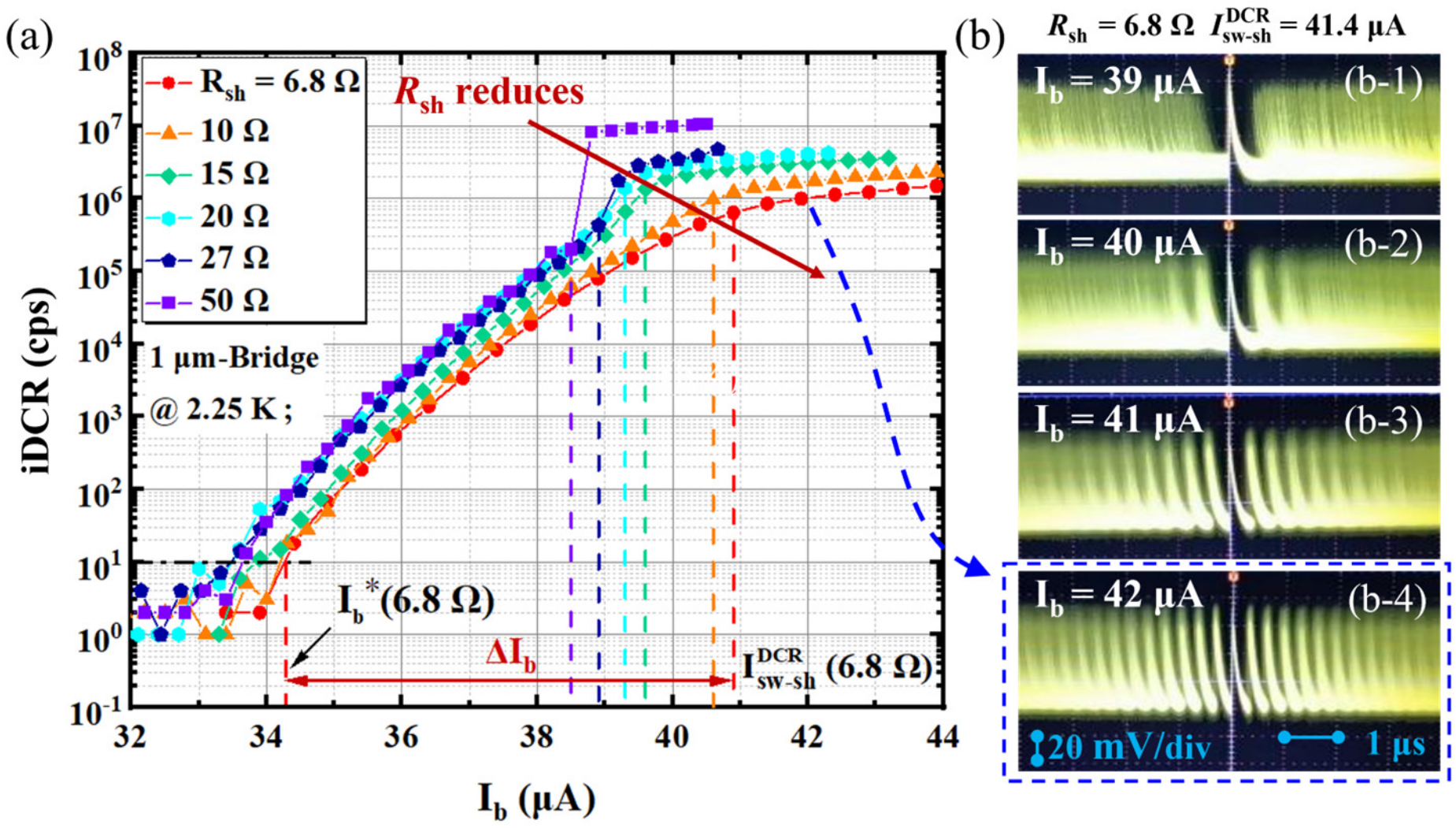

**Figure 3.** (a) iDCR of the 1-μm bridge device as a function of $I_b$, with the vertical axis on a logarithmic scale. The vertical dashed lines indicate the equivalent switching current ($I_{sw-sh}^{DCR}$) corresponding to different $R_{sh}$ values determined from the iDCR curve. The intersection between the iDCR curve and a horizontal dashed line defines the bias current corresponding to a DCR of 10 cps at a certain $R_{sh}$, denoted by $I_b^*(R_{sh})$. (b) Pulse waveforms generated by the dark count near $I_{sw-sh}^{DCR}$, recorded using the oscilloscope, with a shunt resistance of ~ 6.8 Ω and $I_{sw-sh}^{DCR}$ of ~ 41.4 uA. Figure 3(b-4) shows that after exceeding the critical point, an oscillating waveform appeared, e.g., at $I_b$ = 42 μA.

Here we use the intrinsic DCR curves to define the $\Delta I_b$ of SMSPD based on the following considerations: (1) Generally, the SMSPD photo-response curve is affected by many factors such as incident photon energy, input photon intensity, operating temperature, and microstrip geometrical structures. For example, there is a situation that unsaturated photo-response detection exits (such as, the un-irradiated NbN SMSPD operated at 1550 nm), and it could not use the saturation plateau of detection efficiency to evaluate $\Delta I_b$. Besides, the device is likely to latch at high input photon intensity, resulting in an unfair comparison of $\Delta I_b$. (2) To establish a general definition, we chose to use the current dependence of the intrinsic DCR, which is a physical quantity that is less affected by external factors when the detector is operated at a certain temperature and the electrical noises are well eliminated. Of course, the definition of the starting point of $\Delta I_b$ can be adjusted according to one's own needs. The intrinsic DCR of 10 cps in this study is selected according to the acquisition time of the dark count rate in the experiment (1 s in Figure 3, 10 s in Figure 4), in order to better distinguish each DCR curve. However, for the endpoint of $\Delta I_b$, the appearance of the oscillation waveforms could be a good criterion for judgments.

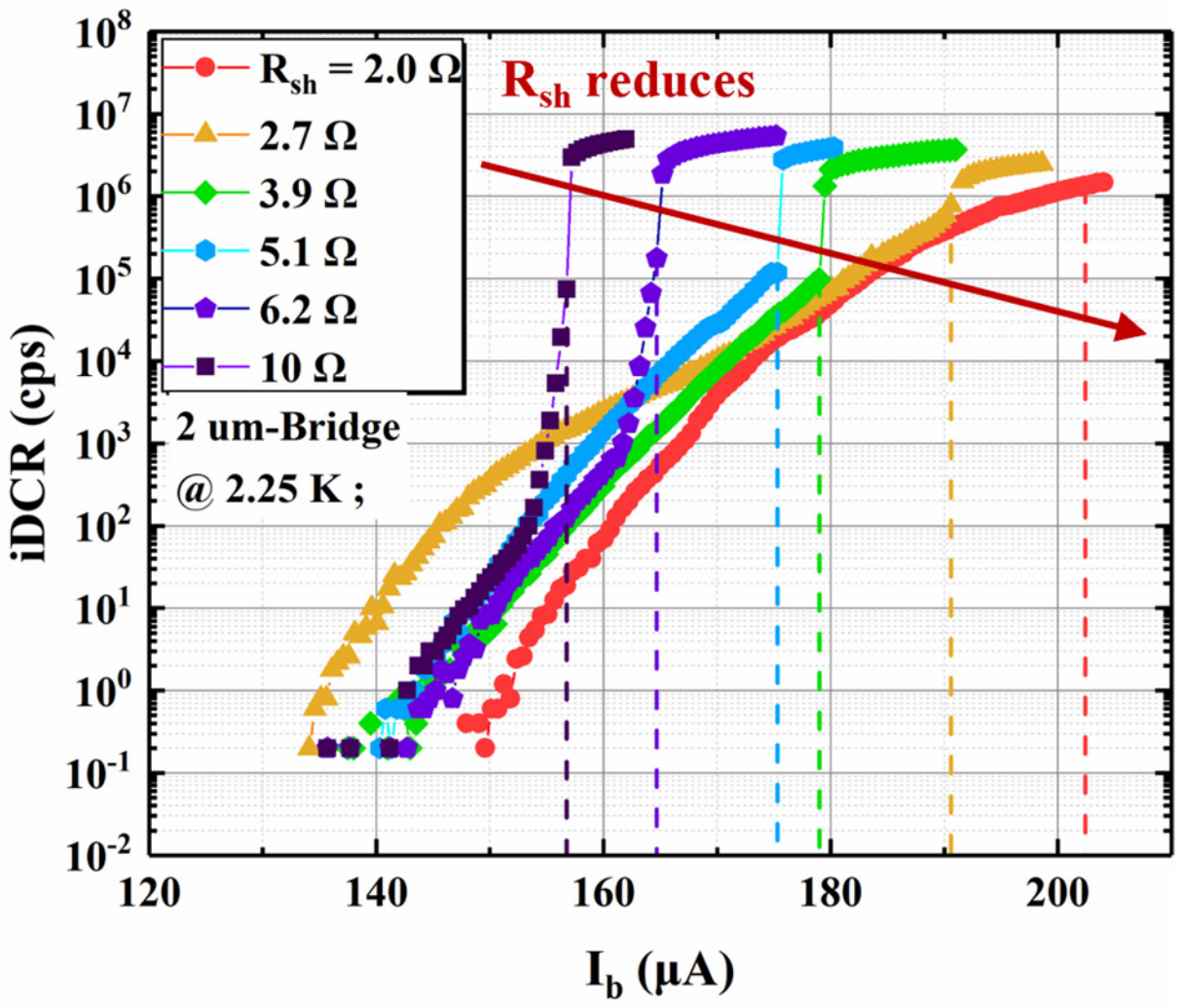

**Figure 4**. iDCR of the 2-μm bridge device as a function of bias current ($I_b$), with the vertical axis on a logarithmic scale. When $R_{sh}$ ≥ 2.7 Ω, a sudden "jump" in dark counts to the oscillation region was observed. The dashed lines indicate the equivalent switching current for each case of $R_{sh}$.

The effect of the shunt resistance on the iDCR behavior is more pronounced in high switching-current devices. As shown in Figure 4, a sudden "jump" in dark counts to the oscillation region was observed for the 2-μm microbridge device shunt with $R_{sh}$ = 2.7–10 Ω. For the 2-μm bridge device, its proper shunt resistance was observed when $R_{sh} \leq 2$ Ω. Therefore, for devices with high critical currents, the choice of the $R_{sh}$ resistance needs to be lower. Notably, when $R_{sh}$ = 2.7 Ω, the DCR was higher than other data in the range of 130–175 μA. This was because, during the measurement, the wrapped shielding aluminum tape was peeled off, and the device was exposed to the black-body radiation remaining in the cryostat. However, this accident did not change the general trend of the effect of the shunt resistance on the DCR. The following analysis excluded the wrong data extracted from this curve so as not to draw wrong conclusions.

The iDCR curves of the 1- and 2-μm bridge devices showed that a suitable shunt resistance is important to extend the width of the normal operating region of the device to improve the IDE. Contrary to the theoretically calculated maximum shunt resistance values obtained in a previous study [25], the present results showed that the theoretical maximum shunt resistance is a prerequisite for preventing latching. However, its value may be unsuitable for achieving a high $\Delta I_b$.

Figure 5(a) summarizes the equivalent switching current defined using two methods as a function of the reciprocal of the shunt resistance ($1/R_{sh}$) for the 1- and 2-μm bridge devices. First, we observed that the equivalent switching-current values determined using these two methods were slightly different. The data in the high-value region of $1/R_{sh}$ could not be determined using the conventional method, owing to the disappearance of the inflection point in the I–V curve. However, they could be obtained in the iDCR curve by monitoring the pulse waveforms using an oscilloscope. Second, a near-linear relation between the equivalent switching current of the same device and the $1/R_{sh}$ was observed, which can be used to predict the $I_{sw-sh}^{DCR}$ in the experiment. Specifically, by measuring two shunt resistance values in advance, the corresponding $I_{sw-sh}^{DCR}$ can be obtained, and the linear relation is used for fitting to predict the $I_{sw-sh}^{DCR}$ for any shunt resistance value on the fitting line.

Notably, the current drift behavior shown in Figure 5(a) induced by the shunt resistance increased the nominal switching current. The maximum enhanced ratio of $I_{sw-sh}^{DCR}/I_{sw}$ for the 1- and 2-μm bridge device was ~ 1.08 and 1.46, respectively. Figure 5(b) shows the $\Delta I_b$ of the two bridges as a function of $1/R_{sh}$, showing that the actual switching current increased as $1/R_{sh}$ increased. We could not continue to perform the measurements when $1/R_{sh} > 500$ ($R_{sh} < 2$ Ω) because of the limitation of the electrical noise of the readout circuit. However, we speculated that $\Delta I_b$ did not linearly increase, like the equivalent switching current, but gradually tended toward a saturated value owing to the constraint of a critical current value in the device. The right axis in Figure 5(b) shows the calculated ratio of $\Delta I_b/I_{sw}$ for each device with the shunt resistance, compared with that of their original switching (without a shunt), distributed within the range of 5%–35%.

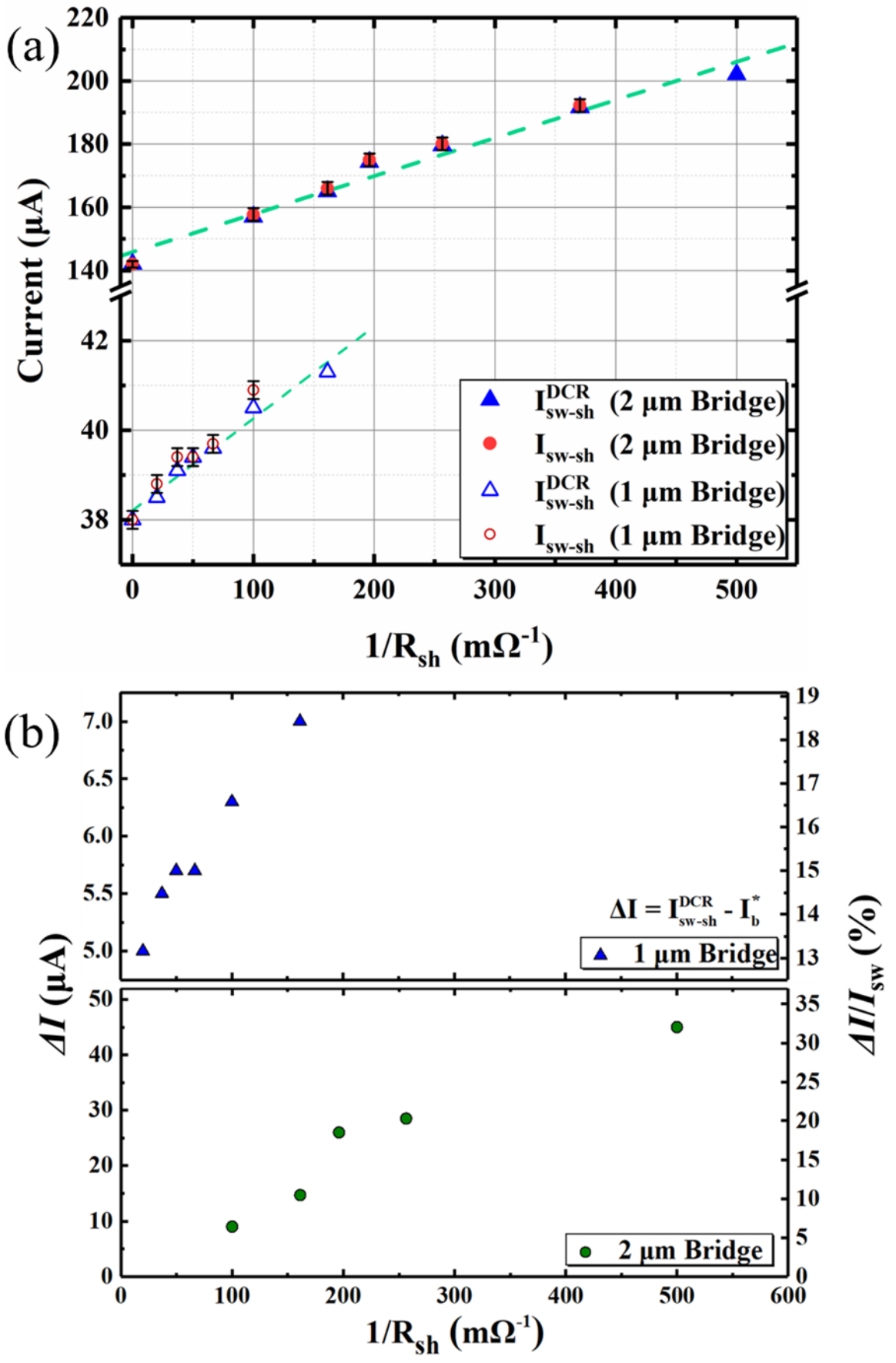

**Figure 5.** (a) Equivalent switching currents of the 1- and 2-μm bridges extracted in two ways as a function of the reciprocal shunt resistance ($1/R_{sh}$). The $I_{sw-sh}^{DCR}$ determined from the iDCR curve is represented by triangular symbols, whereas the $I_{sw-sh}$ extracted from the I–V curve using the conventional method is represented by circular symbols. The dashed lines are an eye guide. (b) Operating bias current range, $\Delta I$, as a function of $1/R_{sh}$, for the 1- and 2-μm bridge devices. The right axis shows the corresponding ratio of $\Delta I/I_{sw}$ calculated for each device in percentage.

### *3.3 Dependence of readout pulse on shunt resistance*

Practically, the shunt resistance cannot be infinitely reduced because it will change the impedance and voltage redistribution of an SMSPD, resulting in changes in the output pulses of the device. Figure 6(a) shows the output pulse waveforms of the 2-μm microbridge at different shunt resistances. As the shunt resistance decreased, the output pulse amplitude ($V_{pulse}$) reduced while the falling edge time ($\tau_{fall}$) of the pulse increased, indicating that the shunt resistance degraded the timing jitter and the count rate performance of the device.

Figures 6(b) and (c) show the measured $V_{pulse}$ and $\tau_{fall}$ as a function of $R_{sh}$ with solid symbols for the 1- and 2-μm microbridge devices, respectively, determined using the oscilloscope. In our readout circuit, owing to the electrical noise and reflections, the minimum trigger level was empirically set to ~ 40 mV for a proper signal readout, which was approximately twice the noise level. To avoid the impact of pulse overshoot on the falling edge, we determined $\tau_{\text{fall}}$ with a criterion of 1/e of the maximum pulse amplitude of the fitted exponential-decay function.

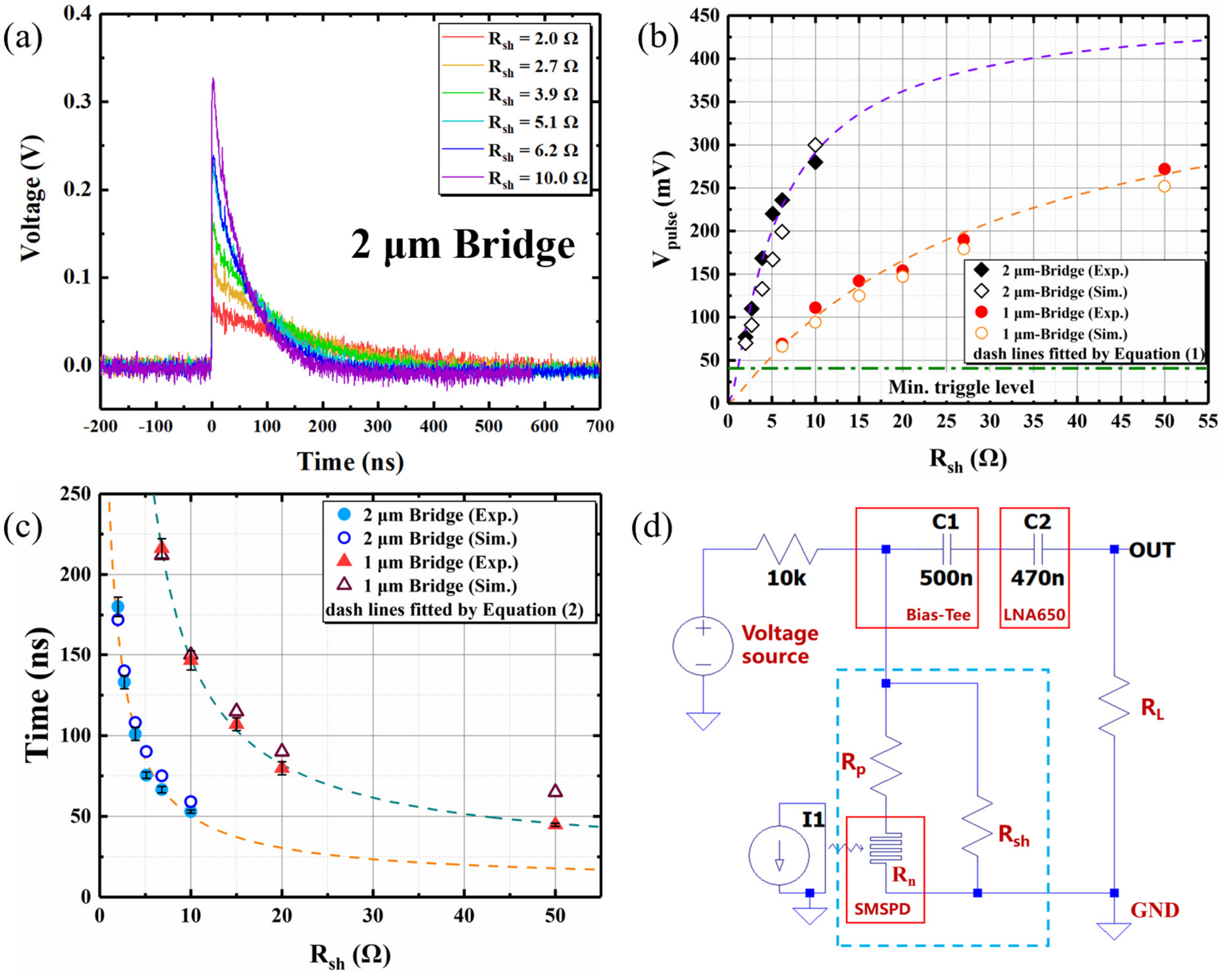

**Figure 6.** (a) Pulse waveforms of the 2-μm bridge device at different shunt resistances recorded using an oscilloscope, with a bias current of $0.9I_{sw-sh}^{DCR}$ ; (2) Pulse amplitude as a function of shunt resistance for the 1- and 2-μm bridge devices; the solid symbols show the experimental data read from an oscilloscope, and the open symbols indicate the simulated data using the LTspice circuit model. The dashed line is the result of fitting using Equation 1. The dotted-dashed line marks the lowest trigger level in the experiment. (c) Pulse decay time as a function of shunt resistance; the solid symbols show the results extracted from the exponential function. The open symbols indicate the results of fitting using LTspice, and the dashed lines show the results of fitting using Equation 2. (d) Schematic diagram of the circuit model used for fitting and simulating the SMSPD shunt with a resistor. The dashed box indicates the package block was placed at cryogenic temperature.

According to Kirchhoff's circuit laws and the schematic diagram shown in Figure 6(d), $V_{\text{pulse}}$ and $\tau_{\text{fall}}$ were expressed as follows, respectively, with the simplified parameters of the bias tee and amplifier:

$$V_{pulse} = G \times I_b \times \left(-\frac{R_p \times R_{sh}}{R_p + R_{sh}} + \frac{(R_p + R_n) \times R_{sh}}{(R_p + R_n) + R_{sh}}\right). \quad (1)$$

$$\tau_{fall} = \frac{L_k}{R_p + R_{sh}//R_L}. \quad (2)$$

Equation 1 describes the amplified voltage pulse at the output port caused by the superconducting–resistive switching of the SMSPD, where $G$ ~ 50 dB (~300 times) is the gain of the amplifier, $I_b$ ~ $0.9I_{sw-sh}^{DCR}$, $R_p$ is the residual resistance of the SMSPD branch, and $R_n$ (~100–200 Ω) represents the resistive region resistance of the SMSPD at 10 K. For Equation 1, the first item in the brackets represents the branch resistance when the device was superconducting, and the second term represents the increasing branch resistance when the device switched into a resistive region. In Equation 2, $R_L$ = 50 Ω is the load resistance of the circuit, and "//" represents the $R_{sh}$ "in parallel" with the $R_L$. Equation 2 indicates that $\tau_{\text{fall}}$ exhibited an inversely proportional

**Table 1.** Predicted minimum and empirical values for selecting a suitable shunt resistor in the microbridge devices. $T_c$ is the critical temperature, and $R_{sq}$ is the sheet resistance of the SMSPDs. The superscript "†" indicates that the devices are presented in this work.

| Sample Type | Dose (ions/cm$^2$) | Width (μm) | Thickness (nm) | $T_c$ (K) | $R_{sq}$(20 K) (Ω/sq) | $I_{sw}$ (μA) w/o $R_{sh}$ @ 2.2 K | $R_{sh}$ (Predicted) | $R_{sh}$ (Empirical) |
|---|---|---|---|---|---|---|---|---|
| Bridge (Unirradiated) | 0 | 1 | 7 | 7.14 | 839 | 100.5 | 2.3 | 5.6 |
| | | 2 | 7 | 7.10 | | 190.1 | 1.1 | 2.0 |
| | | 1 | 8 | 8.07 | 655 | 176.3 | 1.2 | 2.0 |
| | | 2 | 8 | 8.15 | | 355.6 | 0.5 | 1.5 |
| | | 3 | 8 | 8.15 | | 508.4 | 0.4 | 1.0 |
| Bridge† | 1E16 | 2 | 7 | 6.2 | 1005 | 138.0 | 1.4 | 2.0 |
| | 5E16 | 1 | 7 | 5.26 | 1216 | 38.0/39.8 | 4.0 | 6.8 |
| Spiral† | 5E16 | 1 | 7 | 5.41 | 1231 | 35.0 | 4.0 | 6.8 |

decrease with $R_{sh}$.

We fit the experimental data using Equations 1 and 2, shown with dotted lines in Figures 6(b) and (c), respectively. The well-fitting results indicated that the SMSPD behavior can be described by the circuit model shown in Figure 6(d). With Equation 2, the fitting parameter, $L_K$, for the 1- and 2-μm bridge device was ~ 1000 and ~ 500 nH, respectively, while the measured $L_K$ was 790 and 436 nH, determined using a network analyzer (Agilent 4395A).

In addition, we employed the LTspice simulation [36] with the measured $L_K$ to calculate $V_{pulse}$ and $\tau_{fall}$, based on the circuit model shown in Figure 6(d). The open symbols in Figures 6(b) and (c) indicate the simulated results. During the simulation, the minimum simulated shunt resistance was determined by the minimum output pulse amplitude, which should be higher than the lowest trigger level of ~40 mV. Thus, the corresponding minimum shunt resistances for the 1- and 2-μm bridge devices were approximately 4.0 and 1.4 Ω, respectively. The simulated data qualitatively reproduced the trends of the experimental results shown in the figures.

Table 1 shows a summary of the predicted minimum and empirically suitable shunt resistances for the devices in the present experiments. The empirical values were obtained from the experiments, which considered the antilatching, and ensured a relatively high pulse amplitude in the readout system. The data in the table also revealed that the shunt resistance was related to the switching current of the devices, i.e., the higher the switching current, the lower the shunt resistance required. This behavior implies that the proper shunt resistance is temperature-dependent since the critical current is temperature-dependent.

Notably, we also measured the photon-response count rate of the 1-μm Bridge at 0.85 K, shunted with a 6.2-Ω resistance and illuminated with photons of different wavelengths. The photon-response behavior of the microbridge was found to be similar to that of the spiral SMSPD, which will be shown below, and therefore, we will not elaborate on it here.

### *3.4 Improvement of IDE with a shunt resistor*

We measured the double-spiral strip SMSPD in parallel with different shunt resistors in a dilution cryostat at 850 mK. Figure 7(a) shows the IDE plotted against bias current, measured at the shunt resistances of 6.2 and 5.2 Ω. Here, IDE was determined by normalizing the photon-response counts to the asymptote value for the sigmoid function fit. With the lower shunt resistance of 5.2 Ω, the equivalent switching current increased by ~ 2.4 μA. Accordingly, the width of the operating current region increased, and the SMSPD exhibited a saturated detection plateau at a wavelength of 1550 nm.

Thereafter, we adopted a supercontinuum laser source and changed the incident photons of different wavelengths from 400 to 2000 nm with acousto-optic tunable filters.

Figure 7(b) shows the measured results and sigmoidal function fits. We observed that the device still showed weak saturation at 2000 nm. To the best of our knowledge, this is the first time a large area detector fabricated with a reasonable thickness material gets such performances. Moreover, as the wavelength increased beyond 1550 nm, the current increase in the overall curve moving to the right decreased. We defined a threshold detection current, "$I_{IDE}$ = 0.5," at which the current corresponded to an IDE of 0.5 (Figure 7(b)). Figures 7(c) and (d) show the nonlinear wavelength and photon energy dependences of $I_{IDE}$ = 0.5, respectively. The nonlinear behavior of photon energy has been predicted by a previous theoretical study in terms of the maximal detection current [13]. In addition, in the low-photon-energy region (<1.1 eV, or > 1030 nm), the relation of $I_{DE}$ = 0.5 vs. photon energy exhibited near linearity. This implied that the detection mechanism of low-energy photons mainly depends on the ratio of $I_b/I_{sw-sh}^{DCR}$, i.e., it needs to be sufficiently close to $I_{dep}$.

Notably, we should have continued to reduce the shunt resistance to a suitable value; however, unfortunately, our device was degraded by solder contamination during the packaging process. Thus, this part of the study could not be continued. Using the LTspice circuit model and the near-linearity $I_{sw-sh}^{DCR}$– $(1/R_{sh})$ relation, we estimated the proper shunt resistance as ~ 4.0 Ω (corresponding to a simulated pulse amplitude of 40 mV), and the fitted equivalent switching current will reach 57 μA. The saturation of the IDE curve for this device would be improved according to the assumption. In the future, we expect to expand the photon saturation wavelength detection range of SMSPDs using high doses of ion irradiation combined with a suitable shunt resistance.

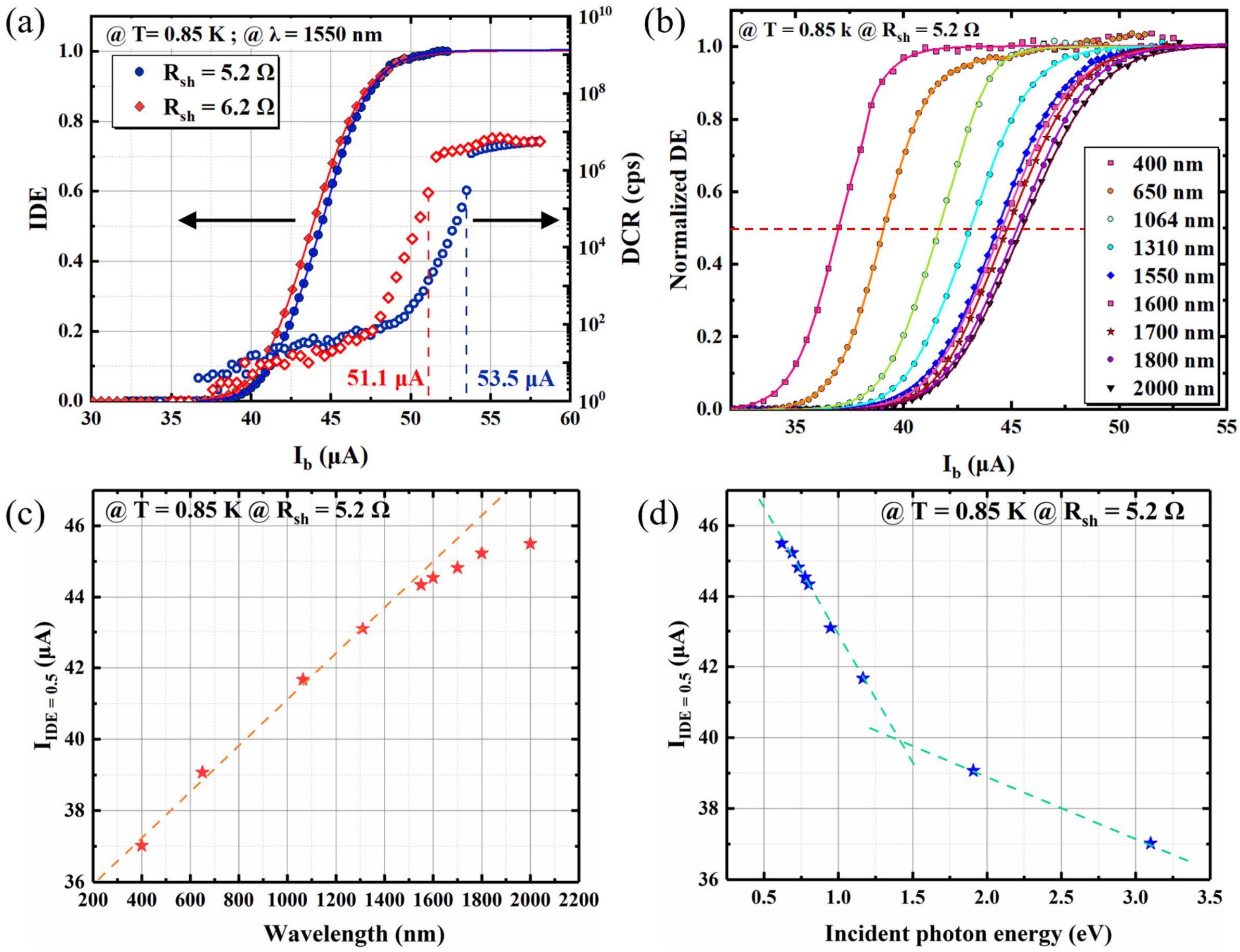

**Figure 7.** (a) IDE and DCR as a function of the bias current for a 1-μm spiral strip SMSPD shunt with two resistors at 6.2 and 5.2 Ω, operated at a temperature of 0.85 K. With a small resistor, the saturation IDE plateau widened. The solid lines are the sigmoid function fits. (b) IDE as a function of the bias current of the same SMSPD shunt with the 5.2 Ω resistor, illuminated under photons of different wavelengths from 400 to 2000 nm. The horizontal dashed line indicates a fractional IDE value equal to 0.5. (c) Detection current corresponding to the IDE of 0.5 (denoted by $I_{IDE=0.5}$) as a function of wavelength. (d) Incident photon-energy dependence of $I_{IDE=0.5}$. The dashed lines are an eye guide.

## 4. Discussion

First, the most direct function of the shunt resistor is to prevent SMSPD latching caused by stable electrothermal feedback [35]. Here, we mainly investigated the relation between the shunt resistance and the operating current range of the device; however, the kinetic inductance of the device can also affect the latching behavior. Latching can be prevented by connecting a sufficiently large inductor in series to the SMSPD [15, 25]. However, the large series inductor can significantly reduce the slew rate on the rising edge of the pulse and increase the recovery time, resulting in increased timing jitter and suppression of the count rate. Second, the choice of the shunt resistor has a trade-off effect. A small shunt resistor can prevent latching and improve the operating current range; however, it can degrade the performance of the device in terms of count rate and timing jitter. It may also cause an impedance mismatch between the detector and readout circuit (typically 50 Ω), thereby distorting the output pulse waveform.

To further study the effect of the combination of series inductance and shunt resistance on the SMSPD performance, we first conducted a study of the variable shunt resistance of the bare microbridge (called Bridge-1#) without series inductance, as shown in Figure 8(a). By reducing the shunt resistance, we found that the bare microbridge can achieve stable operation, and the nominal switching current and bias current range increase monotonically, which was consistent with the behaviors observed in the microbridge with a fixed series inductor in Figure 3, indicating that the effect of shunt resistance on the device is basically the same with or without series inductance, which does not affect our previous conclusions. However, to prevent premature latching of devices without inductance, the suitable shunt resistance required is smaller (≤2.7 Ω, while the equivalent switching current of a series inductance microbridge of ~40 μA was approximately ≤6.8 Ω), resulting in a relatively low output pulse amplitude.

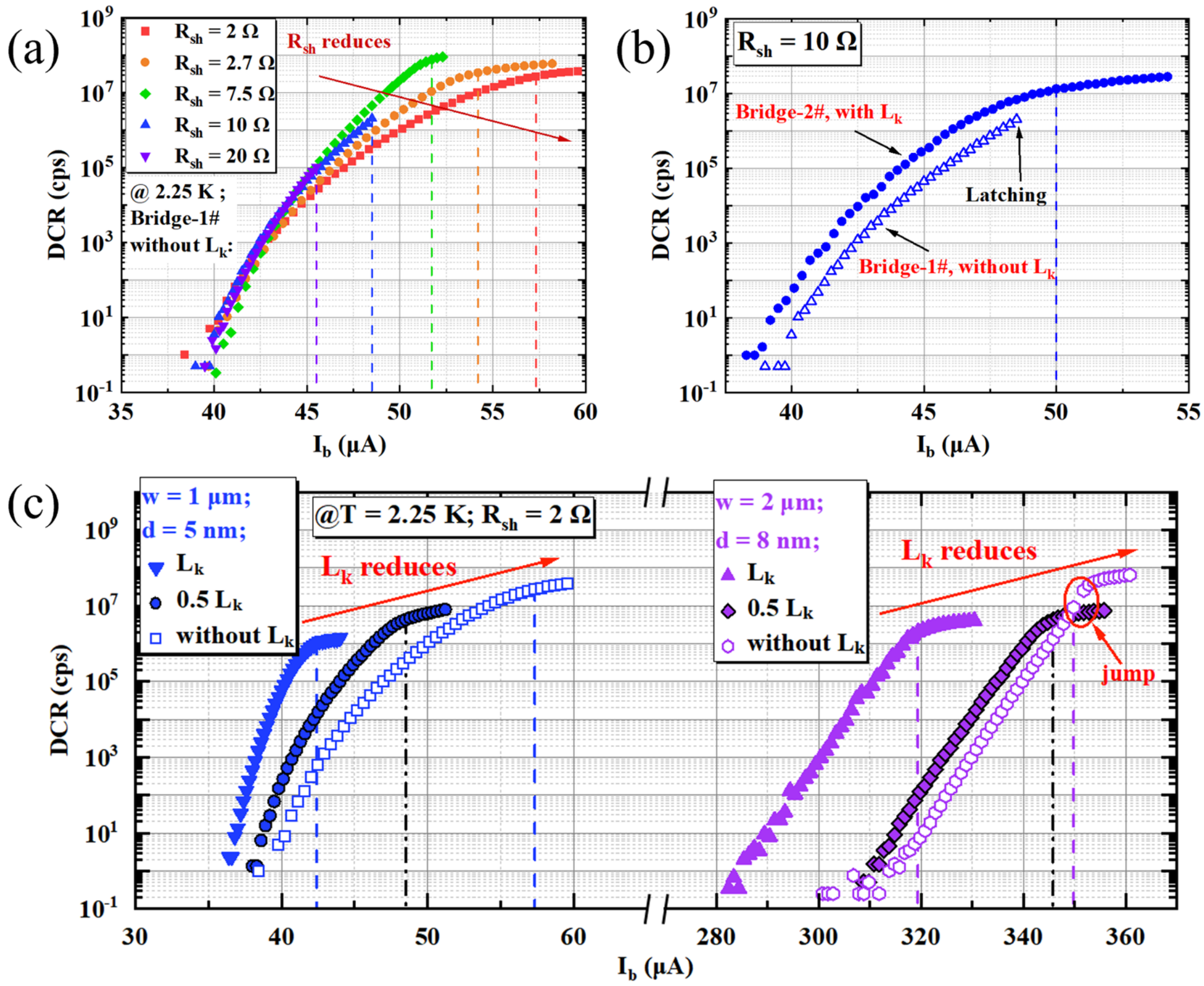

**Figure 8.** The effect of the combination of the shunt resistor and the series inductor on the current-dependent intrinsic DCR curves: (a) A bare microbridge (with no series inductance) connected with different shunt resistors. (b) Comparison of two similar microbridges with and without series inductor when the shunt resistor is relatively large ($R_{sh}$ = 10 Ω). The additional series inductor shows improvement in the stable operating range of the device. (c) Microbridges with different series inductances when the shunt resistance is relatively small ($R_{sh}$ = 2 Ω). As the series inductance decreases, the nominal switching current significantly increases, and the DCR on the oscillation plateau increases significantly.

Next, we compared and measured the intrinsic DCRs of two microbridges with the same geometric structure and very close critical currents (with a difference of <1 μA), where Bridge-2# was connected with series inductance (~ 497 nH) and Bridge-1# without series inductance, as shown in Figure 8(b). With a relatively large shunt resistance ($R_{sh}$ = 10 Ω), premature latching occurred in the Bridge-2# without series inductance at ~48.5 μA. In comparison, the nominal switching current of Bridge-2# with series inductance increased to ~50 μA. The presence of series inductance had a certain improvement on the stability of the SMSPD operation.

Furthermore, we conducted a comparative study of the variation of series inductance for microbridges with different strip widths and thicknesses, which also help evaluate the influence of the difference in current carrying capacity. Here the device configuration design used was the same as that shown in Figure 1(b). We controlled the value of the series inductance by changing the length of the on-chip inductor, i.e., a full length ($L_{KS}$), a half-length ($0.5L_{KS}$), and without $L_{KS}$ (directly wire bonded to the microbridge pad). For the 5-nm-thick, 1-μm-wide on-chip inductor, $L_{KS1}$ ≈ 500 nH; while for the 8-nm-thick, 2-μm-wide on-chip inductor, $L_{KS2}$ ≈ 100 nH. Shunted with a small resistance of 2 Ω, as the inductance decreases, the nominal switching current of the device increases significantly. The oscillation plateau of the dark counts also raised, which reflects a faster pulse recovery time brought about by the decrease in series inductance (also corresponding to the increase in count rates). Since the 2 Ω resistance is small enough relative to the 5-nm-thick microbridge, the trend of dark counts under shunt resistance modulation shows a significant nonlinearity. The results show the negative effect of increasing the inductance on the SMSPDs.

As a comparison, for the high current microbridge with 8-nm thickness, a 2-Ω resistance is relatively large. Although it can prevent latching to some extent, the trend of increasing dark counts was almost linear, indicating that the shunt resistance modulation effect was insignificant. In addition, in the case without series inductance, we observed a "jumping" behavior of dark counts near the plateau, which was clearly observed in Figure 4. This indicates that when the inductance of the SMSPD decreases, the required shunt resistance also needs to be decreased, consistent with the prediction of the electro-thermal feedback mechanism. This also indicates that when the shunt resistance is very small (cannot be infinitely reduced), adding a certain amount of series inductance is necessary to prevent the device's premature latching.

Both inductance and shunt resistance increase timing jitter and suppress count rates. They also caused the shift of the DCR curve along the x-axis, due to the current division caused by the changes of the circuit impedance $Z \propto R + j(\omega L_K - (\omega C)^{-1})$, where $\omega$ is the angular frequency and $C$ is the capacitance. However, they differ in their modulation of the device's electrical-thermal feedback process and the degree of impact: (1) shunt resistance weakens the $I^2R_n$ energy of the hotspot region generated by current heating and modulates the vortex dynamics inside the device, causing the vortex motion to become metastable. For example, the shunt resistance modulation results in nonlinearity in the DCR curves shown in Figure 3, since the measurements were performed with a fixed $L_K$ but varied $R_{sh}$. (2) While Series inductance weakens the continuous heating effect of the current on the device's hotspot by increasing the relaxation time. For example, as shown in Figure 8(c), in the DCR curves with a fixed $R_{sh}$ but varied $L_K$, the shift of the bias current along the X-axis could be attributed to the variation of inductance, while the nonlinear bending in DCR curves was caused by the shunt resistance modulation. This could be a difference between the effect of the shunt resistance and the inductance. (3) Moreover, it is generally believed that the cost of increasing the series inductance on the falling edge time of the output pulse is greater than reducing the shunt resistance [25].

From the above results, it can be seen that the SMSPD showed more complex behaviors when both the shunt resistor and the series inductor are varied. Therefore, we recommend that a normal experimental operation procedure is to use a small shunt resistance under the premise of an acceptable signal-to-noise ratio. If the device does not work stably, an appropriate series inductor needs to be added. However, the final choice still needs to be optimized according to the needs of the actual applications, that is, whether to focus more on the device's higher stable operating area (high intrinsic efficiency) or the device's time jitter and speed performance. Overall, the optimal combination of shunt resistance and series inductance still needs to be quantitatively simulated and measured in the future. Here we also would like to note that the changes in nominal switching current in these measurements could result in some 'illusions on the significant enhancement of switching current'. These illusions may be constrained by the existing measurement or data processing methods (e.g., the scale of the bias current modulated by the circuit impedance variation), which need caution in the SMSPD study.

The shunt behaviors of SMSPDs and SNSPDs share certain similarities, such as the prevention of latching. However, there is a difference in the role of the shunt resistance in these two types of detectors because of the difference in the process of the hotspot or resistive region formation. For example, first, without a shunt resistor, NbN SMSPDs generally cannot function. Second, the detection mechanism of SMSPDs makes them extremely dependent on a high ratio of $I_b/I_{sw-sh}^{DCR}$ to obtain high IDE, particularly when the photon sensitivity of the material is not sufficiently high. Thus, SMSPDs typically need to be biased at a high $I_{sw-sh}$ ratio (~95% $I_{sw-sh}^{DCR}$ in Figure 7, IDE ~100%). The widening of the operating current range (usually 5%–35% of the $I_{sw}$ without a shunt) induced by a suitable shunt resistance can improve the IDE and reduce the iDCR. However, the improvement of the IDE through the shunt resistance is still limited. A high ratio of $I_b/I_{sw-sh}^{DCR}$ is also accompanied by a high iDCR penalty. Thus, the enhancement in the photon sensitivity of the thin-film material is the primary determining factor for improving the IDE. Third, since the current carried by SMSPDs is high and the resistance of SMSPD is relatively low, to realize the unstable electrothermal feedback, the corresponding shunt resistance needs to be low to reduce the heat generation in the strip. Contrarily, for a typical SNSPD, the switching current is usually below 40 μA, and the shunt resistor is generally in the range of 25–50 Ω. The triggering of the pulse response in the SNSPD is mainly through the formation of the hotspot across the nanowire, which has a good tolerance to the ratio of $I_b/I_{sw}$ because of a narrow strip width. Thus, when the ratio of $I_b/I_{sw}$ reaches a certain level (generally > 80% $I_{sw}$) [37], the SNSPD can generate a high probability of photon detection (e.g., IDE ~100% for NbN SNSPD). Thus, the IDE improvement induced by the shunt resistor to the SNSPD is insignificant. The role of the shunt resistor in SNSPDs is to prevent device latching and improve the operation stability of the detector.

## 5. Conclusions

We systematically studied the effects of shunt resistance on the electrical behaviors of the NbN microbridge SMSPD using I–V curves, iDCR curves, and pulse waveforms. We observed that the I–V curve exhibited three regions with different slopes, which changed with the shunt resistance. The iDCR curve of the SMSPD with the shunt resistor exhibited a logarithm nonlinearity as a function of bias current. When the shunt resistance was unsuitable, we observed a "sudden jump" or "inflection point" in the iDCR curve before the detector switched into a "relaxation oscillation region." The iDCR exhibited a slowly rising plateau in the oscillation region, which decreased with a decrease in the shunt resistance. Interestingly, when the shunt resistance was suitable (below a certain value), the inflection point in the iDCR curve gradually disappeared. This phenomenon could be due to the stable vortex dynamics in the strip modulated by the shunt resistor.

We analyzed the variation in the equivalent switching current at different shunt resistances in two approaches (defined from the I–V and iDCR curves) and studied the operating current range through the iDCR curve. We confirmed that the choice of shunt resistance has a trade-off effect. On the one hand, a small shunt resistance can increase the operating current range of the detector, thereby improving the IDE of the device and reducing the iDCR. On the other hand, it can reduce the pulse amplitude and increase the pulse decay time, thereby degrading the timing jitter and count rate performance of the SMSPD. In addition, the selection of the shunt resistor is related to the current-carrying capacity of the strip, i.e., the higher the critical current, the smaller the shunt resistance required. We observed that the experiment data can be well-fitted by the equations deduced from the circuit model of the SMSPD with a shunt resistor. We verified this model with the LTspice simulation, which reproduced the trends of the measured data. Combining the near-linearity $I_{sw-sh}^{DCR}$ –($1/R_{\mathrm{sh}}$) relation with the LTspice simulation, we qualitatively predicted the electrical behaviors of the SMSPD with the shunt resistor in terms of the equivalent switching current, pulse amplitude, and pulse decay time. We also provided empirical recommendations for the selection of shunt resistors.

As a demonstration, we characterized the detection performance of a double-spiral SMSPD with a strip width of 1 μm and an active area of 50 μm; the SMSPD was irradiated with helium ions and shunted with different resistances. Shunted with a small resistance of 5.2 Ω and operated at 0.85 K, the SMSPD exhibited an improved IDE at 1550 nm and weak saturated IDE at a wavelength up to 2000 nm. We observed a nonlinear relation between detection current and photon energy. This study provides useful information for selecting a proper shunt resistor, thereby deepening the understanding of the SMSPD latching mechanism and extending the detection wavelength of SMSPDs.

There are still some issues that need to be addressed in the future. These include finding the optimal combination of shunt resistance and inductance, exploring the limitations of shunt resistance on the intrinsic DCR of SMSPD (which is a concern for dark matter detection), and studying the vortex dynamics and simulations of shunted SMSPDs.

**Acknowledgments**

This work is supported by the National Natural Science Foundation of China (NSFC, Grant No. 61971409), the National Key R&D Program of China (Grants No. 2017YFA0304000), and the Science and Technology Commission of Shanghai Municipality (Grant No. 18511110202, and No. 2019SHZDZX01). W. -J. Zhang is supported by the Youth Innovation Promotion Association (No. 2019238), Chinese Academy of Sciences.

**Author contributions:** W.-J.Z. and Y.-Z.W. conceived and designed the experiments. Y.-Z.W. and G.-Z.X. fabricated the samples, carried out the experiment, and collected the data. Y.-Z.W. and W.-J.Z. analyzed the data and prepared the manuscript. All authors discussed the results and reviewed the manuscript.

**Competing interests:** The authors declare no competing interests.

**Data and materials availability:** Any data that support the findings of this study are included within the article.